\documentclass[final,3p,times]{elsarticle}

\usepackage{lineno,hyperref}
\hypersetup{
colorlinks=true,
citecolor=blue
}
\modulolinenumbers[5]
\journal{Journal of Informetrics}
\biboptions{authoryear}

\usepackage{amssymb}
\usepackage{amsmath}
\usepackage{amsfonts}
\usepackage{graphicx}
\usepackage{dcolumn}
\usepackage{bm}
\usepackage[english]{babel}
\usepackage[utf8]{inputenc}
\usepackage{color}
\usepackage{booktabs}
\usepackage{times}
\usepackage{tablefootnote}
\usepackage{multirow}

\graphicspath{{.}{./Figures/}}

\begin{document}

\begin{frontmatter}

\title{Unraveling the dynamics of growth, aging and inflation for
citations to scientific articles from specific research fields}

\author[vuw]{K.~W. Higham}
\author[vuw]{M. Governale}
\author[motu,qut]{A.~B. Jaffe}
\author[vuw]{U. Z\"ulicke\corref{cor1}}
\ead{uli.zuelicke@vuw.ac.nz}

\cortext[cor1]{Corresponding author \hfill\includegraphics[height=0.5cm]{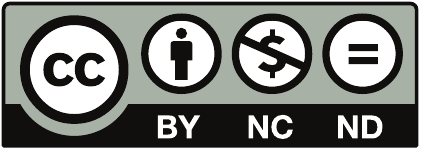}}

\address[vuw]{Te P{\=u}naha Matatini, School of Chemical and Physical
Sciences, Victoria University of Wellington, PO Box 600, Wellington
6140, New Zealand}

\address[motu]{Te P{\=u}naha Matatini, Motu Economic and Public
Policy Research, PO Box 24390, Wellington 6142, New Zealand}

\address[qut]{QUT Business School, Queensland University of
Technology, GPO Box 2434, Brisbane, QLD 4001, Australia}

\begin{abstract}

We analyze the time evolution of citations acquired by articles from
journals of the American Physical Society (PRA, PRB, PRC, PRD, PRE
and PRL). The observed change over time in the number of papers
published in each journal is considered an exogenously caused
variation in citability that is accounted for by a normalization.
The appropriately inflation-adjusted citation rates are found to be
separable into a preferential-attachment-type growth kernel and a
purely obsolescence-related (i.e., monotonously decreasing as a
function of time since publication) aging function. Variations in
the empirically extracted parameters of the growth kernels and aging
functions associated with different journals point to
research-field-specific characteristics of citation intensity and
knowledge flow. Comparison with analogous results for the citation
dynamics of technology-disaggregated cohorts of patents provides
deeper insight into the basic principles of information propagation
as indicated by citing behavior.

\end{abstract}


\end{frontmatter}

\section{Introduction}

Peer-reviewed publications in scientific journals and patents issued
by the national patent offices both serve to codify and document
knowledge advances. To delineate clearly the reported scientific
(technological) progress that has been achieved by the authors
(inventors), citations to prior work (art) are necessary. While the
detailed mechanisms and motivations governing the use of citations to
scientific articles~\citep{garfield1963,garfield2006,bornmann2008}
are generally different from those applying to
patents~\citep{jaffe2000,hall2002,cotropia2013,jaffe2017}, all citing
behavior is widely believed to be indicative of, at least some kind
of, knowledge flow or information transfer. Furthermore, both for
scientific articles and patents, citations are considered to be a
(more or less noisy) proxy measure of impact~\citep{griliches1990,
hall2005,vonWartburg2005,garfield2006,lane2010}. This has motivated
the quantitative study of citations, especially their distributions
across suitably defined cohorts~\citep{dSPrice1965,seglen1992,
redner1998,redner2005,valverde2007,radicchi2008,stringer2010,
vieira2010,radicchi2011,waltman2012,golosovsky2017a,sheridan2017}, as
well as the dynamics of how citations are acquired over
time~\citep{dSPrice1976,avramescu1979,glanzel2004,redner2005,
simkin2007,csardi2007,valverde2007,golosovsky2012,scharnhorst2012,
wang2013,parolo2015,colavizza2016,pan2016,golosovsky2017,higham2017,
yin2017}. The ultimate goal of such investigation is the
establishment of a basic generative model that captures the
fundamental mechanisms governing citation dynamics and can thus 
reproduce the empirically observed time evolution and general 
statistical properties of citation accrual. Ideally, a properly 
validated model would be applicable to inform rational science and 
innovation policies~\citep{lane2010}. 

Recent progress towards realistic, and potentially predictive,
descriptions of citation dynamics~\citep{redner2005,csardi2007,
valverde2007,golosovsky2012,wang2013,pan2016,golosovsky2017,
higham2017} has capitalized on advances in complex-network
theory~\citep{albert2002,dorogovtsev2002,newman2003}. In particular,
the concept of preferential attachment (PA)~\citep{barabasi1999,
dorogovtsev2000,krapivsky2001} governing the rate at which citations
are distributed has been very influential almost from the
beginning~\citep{dSPrice1976}. However, the fruitful application of
PA to understand citation behavior is predicated on the understanding
of two other basic temporal influences: obsolescence and overall
growth of research fields. Here we understand \textit{obsolescence\/}
to be reflected in the tendency for the citation rate to articles or
patents to decay over time because of their reduced relevance for
ongoing knowledge generation. Acting in parallel to the basic trend
towards obsolescence, the overall growth of research fields provides
another important mechanism that influences the rate at which
citations are acquired. Empirical studies have observed a steady
increase over time in the production of scientific
articles~\citep{dSPrice1965,sinatra2015} and
patents~\citep{hall2002}. As every article and patent will generally
have to cite the knowledge stock that is current at the time of their
creation, an increase in article and patent production will likely
lead to an increase in the rate at which prior work is cited. The
need for a careful disentangling of obsolescence and \textit{citation
inflation\/} due to growth was discussed early on, both for
scientific articles~\citep{egghe2000} and patents~\citep{hall2002}.
The most widely adopted method to address growth consists of
introducing normalization factors based on citation
counts~\citep{radicchi2008,radicchi2011,wang2013,yin2017}, which is
partly a result of the desire to find robust bibliometric impact
measures for individual authors or institutions.

Here our motivation is different. We are interested in characterizing
the \textit{intrinsic\/} dynamics of knowledge generation and
propagation that can be revealed by citation behavior if purely
exogenous factors such as changes in article and patent productivity
are appropriately accounted for. Our approach is inspired by its
success in the context of patent-citation dynamics~\citep{higham2017}
and also a recent study~\citep{subelj2017} where normalization by the
number of articles published per year led to the observation of
universal citation distributions for a large body of articles from
physics and computer science, respectively. Furthermore, exponential
growth was used as one ingredient in a successful network-model
simulation of citations to scientific articles~\citep{wang2013}.
Similar to our previous work on patents~\citep{higham2017}, we
analyze citations within different research fields/subfields of
physics as defined by the scope of individual journals published by
the American Physical Society. The obtained journal-specific
characteristics for the PA mechanism and obsolescence function are
indicative of special features associated with knowledge generation
and propagation in different physics-researcher communities.

\section{Data, methods and results}

The bibliometric and citation data set used in our work is provided
by the American Physical Society (APS) and, in its entirety, consists
of article metadata and citation pairs dating back to
1893~\citep{APSJournals}. The subset of this data set that we focus
on here are the cohorts of articles published in the year
2000\footnote{Our choice of this particular year constitutes a
compromise between us capitalising on the increase over time in
publication rates to maximise the article-cohort sizes while, at the
same time, keeping a large-enough time window for articles to garner
citations and facilitate the reliable observation of citation growth
and obsolescence.} in the research-field-specific APS journals
Physical Review A, B, C, D, and E (from this point onwards
abbreviated as PRA, PRB, etc.), as well as the APS's
multidisciplinary-physics letters journal Physical Review Letters
(PRL). Citation rates are measured using all citation pairs whereby
the cited article in one of these cohorts is linked to a citing
article published in the years 2000--2015 in any APS journal.
Table~\ref{statstable} provides an overview of the journal-specific
article cohorts, with citation-number totals and other relevant
citation-related statistical information. For all of the specialized
journals (i.e., PRA--E), the fraction of citations originating from
articles published in the same journal is quite high, justifying our
approach to use these journals to be representative of different
research fields. As expected, this is not the case for the
multi-disciplinary letters journal PRL, which we include in our
study as a benchmark for useful comparison.

\begin{table*}
\caption{\label{statstable}%
Summary statistics for the APS-journal-article citation data set.
Article cohorts comprise all articles published in a given journal in
the year 2000. We analyse citations to these from other APS-journal
articles published up until the end of the year 2015. The fraction of
journal self-citations quantifies the number of citations originating
from articles in the same journal where the cited article was
published. In addition to listing the total number of citations
accrued by each cohort, we also give the total of inflation-adjusted
citations that is obtained by summing the citation counts that have
been scaled to control for variations in citability due to changes in
the numbers of articles published at different times. For reference,
the mean and median numbers of inflation-adjusted citations per
article are also provided for each APS-journal cohort.}
\vspace{0.2cm}
\centering{\begin{tabular}{|c|cccccc|}
\hline Journal & PRA & PRB & PRC & PRD & PRE & PRL \\ \hline
Number of articles published in 2000 & 1,458 & 4,994 & 863 & 2,049 &
2,255 & 3,123 \\ \hline
Total number of citations accrued by cohort & 17,005 & 49,417 & 7,959
& 26,072 & 14,675 & 70,876 \\ \hline
Fraction of journal self-citations & 0.74 & 0.80 & 0.82 & 0.91 & 0.70
& 0.23 \\ \hline
Total number of inflation-adjusted citations & 12,927 & 44,401 &
7,389 & 19,776 & 11,423 & 59,083 \\ \hline
Mean inflation-adjusted citations per article & 8.87 & 8.89 & 8.56 &
9.65 & 5.07 & 18.92 \\ \hline
Median inflation-adjusted citations per article & 4.21 & 4.71 & 5.01 
& 4.59 & 2.81 & 9.99 \\ \hline
\end{tabular}}
\end{table*}

\begin{figure}[t]
\centering{\includegraphics[width=0.7\textwidth]{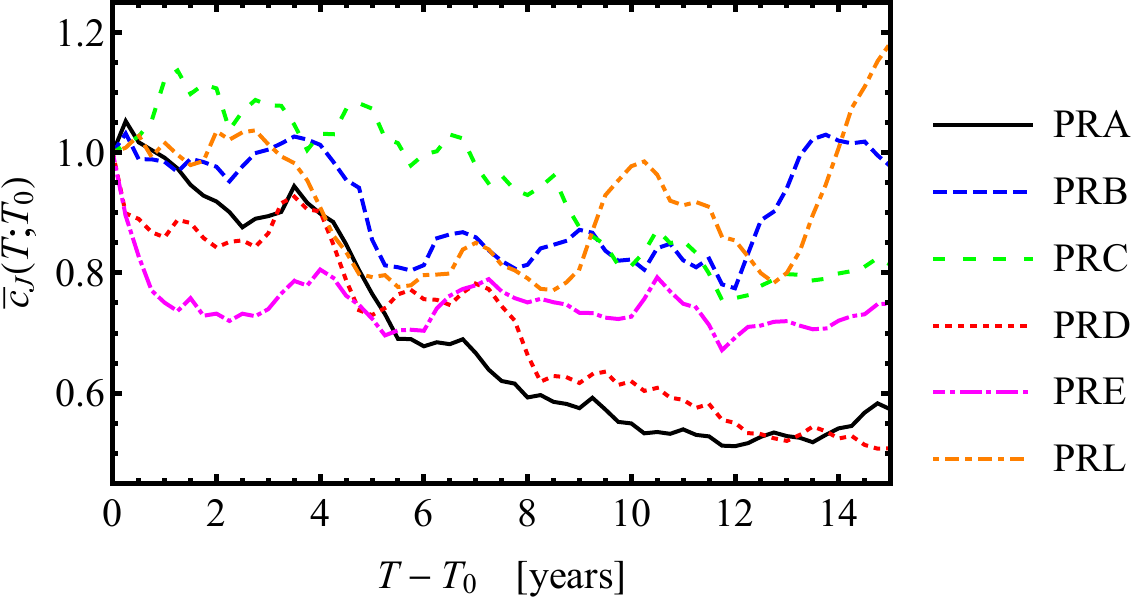}}
\vspace{-0.2cm}
\caption{\label{fig:cite_value}%
Variation of inflation-adjusted citation values for individual APS
journals over the 15-year period starting in 2000. We calculated the
citation-value parameter $c_J(T; T_0, \Delta T)$ defined in
Eq.~(\ref{eq:citeVal}) for $T_0=1$~January 2000 and $\Delta T =
3$~months. Hence, the value of a citation made in a journal article
published at time $T$ corresponds to the ratio of the number of
articles published in that journal in the first quarter of the year
2000 to the number of articles published in that same journal in the
3-month period starting at $T$. To smooth short-term temporal
fluctuations, we plot here the average $\bar c_J(T; T_0) \equiv
\left[ \sum_{m=-2}^{2} c_J(T + m\,\Delta T; T_0, \Delta T)\, \Theta
(T - |m|\, \Delta T - T_0)\right]/\sum_{m=-2}^2 \Theta(T-|m|\,\Delta
T - T_0)$, with $\Theta(x)$ denoting the Heaviside step function.
Most journals exhibit a long-term trend of citation inflation due to
the overall increasing rate at which articles are published. PRL and
PRB are notable exceptions due to recent changes in their editorial
policies~\citep{meystre2013,molenkamp2013}.}
\end{figure}

To be able to separate the various mechanisms that together determine
the rate at which citations are gained by scientific articles, we
first devise a procedure to account for temporal variations in
citeability arising from purely exogenous driving forces. Both the
number of articles produced and the average number of citations made
by each article vary over time~\citep{radicchi2011,wang2013,
sinatra2015,pan2016}. In the long term, the combined effect
of these factors causes a citation inflation that can mask the trend
of obsolescence. In this work, we consider the changing rate at which
articles are published an exogenous factor, as such changes in
research productivity can be expected to be largely determined by
the availability of resources, general policy decisions, or other
influences that do not reflect the utility of prior knowledge. In
contrast, any change in the average number of citations made by each
article is indicative of the need to cite more or less of the
currently relevant knowledge and, thus, is intrinsic to the
information ecosystem. Based on this philosophy, we `deflate’ the
value of incoming citations to each journal-specific cohort in each
3-month period such that the value of a citation to a particular
article in a particular quarter is scaled by the ratio of the total
number of articles published in the citing journal in the first
quarter of the year 2000 to the total number of articles published in
the same journal in the quarter in question. That is, if there were
twice as many PRE articles published in the third quarter of the year
2010 as there were in the first quarter of the year 2000, then
citations given by articles published in the former period would be
given a value of 0.5 to reflect the higher chance of attaining a
citation from that journal due to extrinsic growth in
article-publication rates. Figure~\ref{fig:cite_value} illustrates
the time evolution of the thus-defined value $c_J$ for citations
originating from different APS journals $J\in \{\mbox{PRA},
\mbox{PRB}, \mbox{PRC}, \mbox{PRD}, \mbox{PRE}, \mbox{PRL}\}$, whose
explicit mathematical expression is given here also for greater
clarity;
\begin{equation}\label{eq:citeVal}
c_J(T; T_0, \Delta T) = \frac{N_J(T_0, \Delta T)}{N_J(T, \Delta T)}
\quad ,
\end{equation}
where $N_J(T, \Delta T)$ is the number of articles published in
journal $J$ in the time interval $[T, T+\Delta T]$. 

After citations have been inflation-adjusted according to the
procedure described above, all citations to articles in our cohort
are assigned a time $t$ corresponding to the time lag between the
publication of the cited article and that of the citing article. In
order to model the dependence of the citation rate as a function of
both time $t$ and the number $k$ of accrued citations, we bin
citations by time of arrival where each bin has a range $(t,t+
\Delta t]$. We therefore observe two time series: the number of
citations each article $i$ has accrued by time $t$, denoted by $k_i
(t)$, and the number of citations each article gains in the next
period $(t,t+\Delta t]$, denoted as $\Delta k_i (t)$. Our further
analysis will be based on the assumption that the rate $\lambda(t)$
at which individual articles gain citations is a function of both $k$
and $t$,
\begin{equation}
\lambda(t) = \bar\lambda(k(t), t) \quad .
\end{equation}
As an empirical measure for $\bar\lambda(k, t)$, we use the average
citation rate for the group of articles with $k$ citations in the
interval $(t,t+\Delta t]$, i.e., $\bar{\lambda} (k(t),t) \approx
\overline{\Delta k_i (t+\Delta t)}/\Delta t$. In this manner, we
obtain a matrix of citation rates for each journal with every entry
corresponding to the average rate of citation to the group of
articles published in that journal in the year 2000 with $k$
citations at time $t$. We therefore place each article in a
particular bin based on its accrued citations, $k_i$, at the end of
each time period. As described above, the binning in the
time dimension is simply a linear scale where each bin has the width
$\Delta t$. In the $k$ dimension, logarithmic binning is used. This
means each `$k$-bin’ has the same width on a logarithmic scale, which
turns out to be appropriate for the observed functional form for the
$k$ dependence of $\bar\lambda(k,t)$. In the implementation of this
binning, we first introduce a threshold set at the 99th-percentile
level of accrued citations by the end of the year 2015 (by journal).
Once an article gains more than this number of citations, it is
excluded from our measurements. This is done because there are not
enough data in each bin above this threshold to measure citation
rates accurately, and the large variance introduced by these data
points would negatively affect the measurement of our model
parameters.

\begin{figure}[t]
\includegraphics[width=0.48\textwidth]{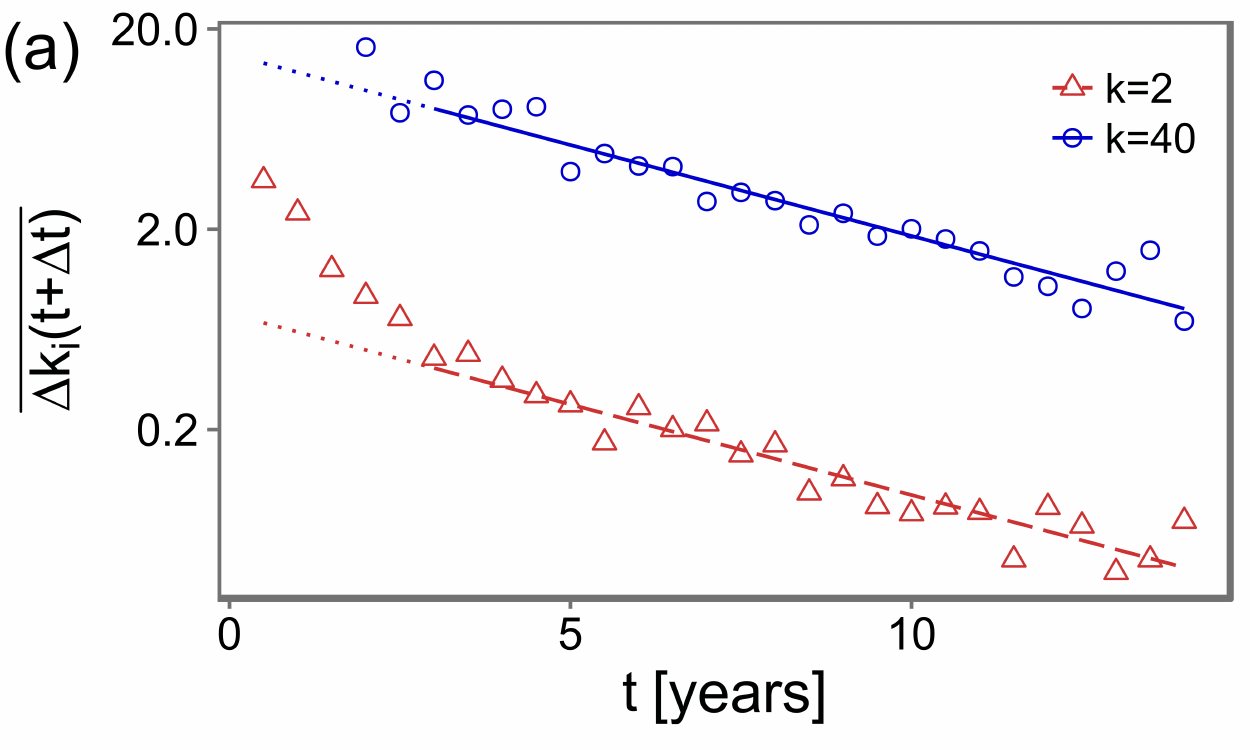}\hfill
\includegraphics[width=0.48\textwidth]{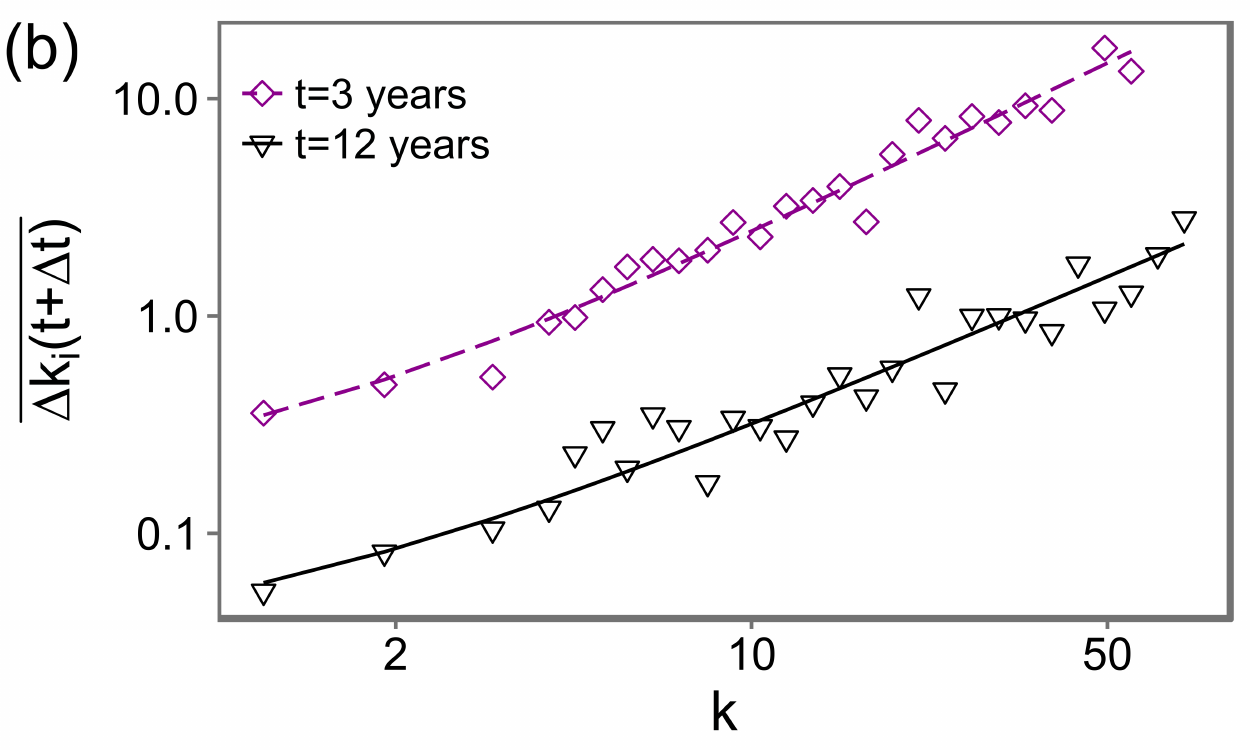}
\vspace{-0.4cm}
\caption{\label{fig:fits}%
Determining the time ($t$) and citation-number ($k$) dependences of
the citation rate $\bar\lambda(k, t)$, using the average number of
additional citations gained in the time interval $(t, t+\Delta t]$,
denoted by $\overline{\Delta k_i (t+\Delta t)}$, as its empirical
measure. Data shown here are for articles published in PRD in the
year 2000. (a)~Symbols show the time dependence of $\overline{\Delta
k_i (t+\Delta t)}$, with $\Delta t = 6$~months, for articles from
bins with logarithmic-scale midpoints at $k=2$ and $k=40$. The curves
are fits of the aging function from Eq.~(\ref{eq:ageing}) to the
data. (b)~Symbols show the dependence of $\overline{\Delta k_i (t+
\Delta t)}$, with $\Delta t = 1$~year, on $k$ at fixed times $t=
3$~years and 12 years. Curves show fits of the PA expression from
Eq.~(\ref{eq:prefatt}) with $f_0=1$ to the empirical citation rate.}
\end{figure}

The simultaneous influences of knowledge-diffusion-driven growth and
obsolescence-related decay on the citation rate can be captured by
postulating the functional form~\citep{dorogovtsev2000a,zhu2003,
csardi2007,valverde2007,golosovsky2012}
\begin{equation}\label{eq:sepaCit}
\bar\lambda (k, t) = A(t)\, f(k) \quad ,
\end{equation}
where $A(t)$ is a purely time-dependent aging function, and the
growth kernel $f(k)$ embodies the PA mechanism. That the rate at
which individual articles gain citations is indeed of the separable
form (\ref{eq:sepaCit}) is nontrivial and needs to be tested. To this
end, we have fitted the observed $t$ and $k$ dependences of the
citation rate for the article cohorts from a given journal and find
that the observations are best described by the functional forms
\begin{subequations}
\begin{eqnarray}\label{eq:ageing}
A(t) &=& A_0\, \exp\left(-\frac{t}{\tau}\right) \quad \mbox {for
$t \ge t_0\approx \tau/2$} \quad , \\[0.1cm] \label{eq:prefatt}
f(k) &=& k^\alpha + f_0 \quad .
\end{eqnarray}
\end{subequations}
Figure~\ref{fig:fits} shows examples of the performed
fits\footnote{All fitting is completed using nonlinear logarithmic
regressions from which parameters and their variances are
determined.}. A possible alternative form of $f(k)$ is discussed in
\ref{sec:appA}.

\begin{figure}[t]
\includegraphics[width=0.48\textwidth]{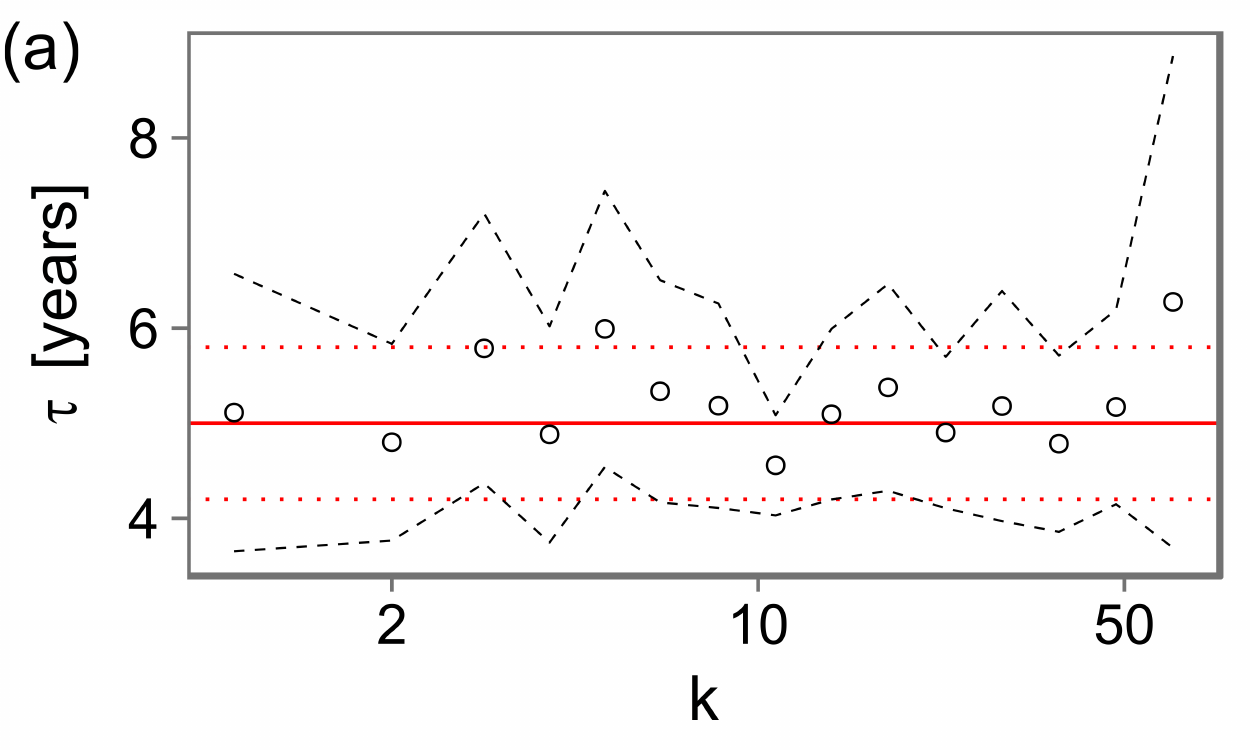}\hfill
\includegraphics[width=0.48\textwidth]{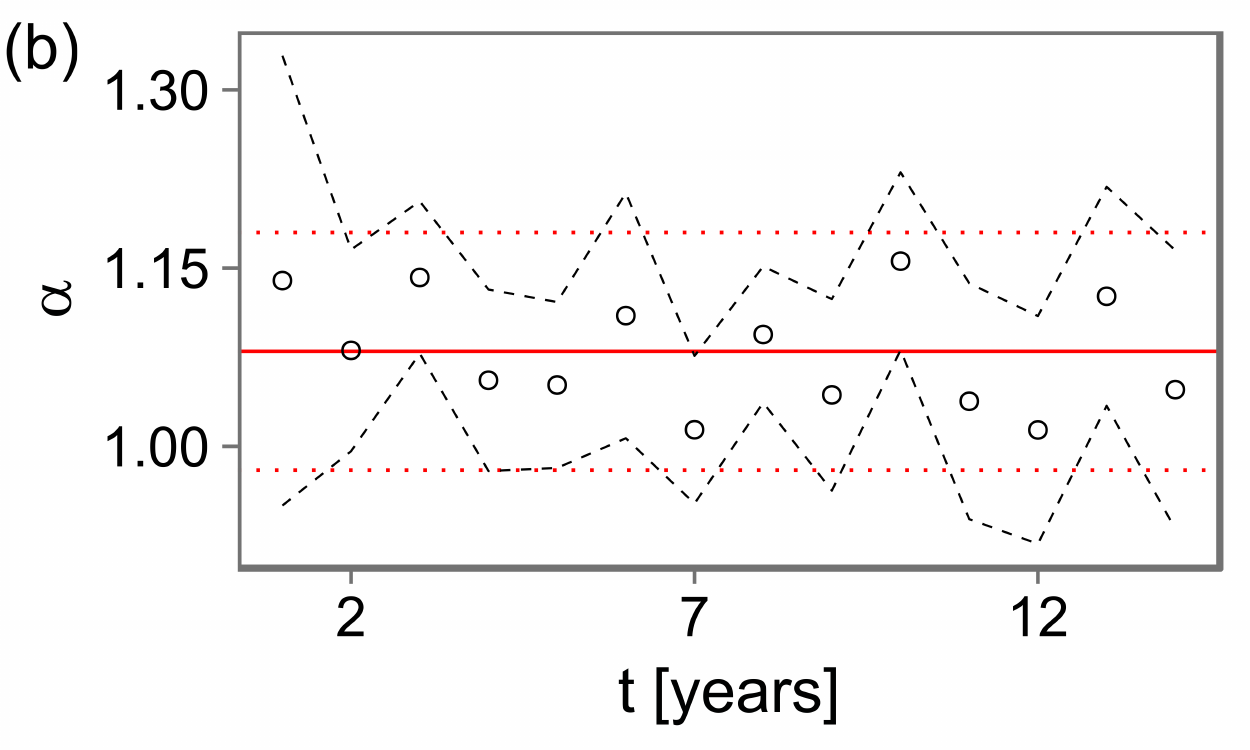}
\vspace{-0.4cm}
\caption{\label{fig:parameters}%
Demonstrating separability of the citation rate into a purely
$t$-dependent aging function and a $k$-dependent PA growth kernel, as
expressed in Eq.~(\ref{eq:sepaCit}). Data shown pertain to articles
published in PRD in the year 2000. (a)~Values of the obsolescence
time $\tau$ extracted from fits of the empirical citation rate to the
functional form $A(t)$ from Eq.~(\ref{eq:ageing}) for different fixed
$k$. (b)~Values for the exponent $\alpha$ derived from fits of the
empirical citation rate to the form (\ref{eq:prefatt}) for the PA
growth kernel $f(k)$, assuming $f_0=1$. Circles are the fitted
parameter values, solid lines indicate their weighted averages, and
the black dashed (red dotted) curves show the 95\% confidence
intervals for fit-parameter values (their weighted averages).}
\end{figure}

If the separability of the citation rate into $t$ and $k$-dependent
factors according to Eq.~(\ref{eq:sepaCit}) holds, then the
empirically extracted values of $\alpha$ and $f_0$ should be
independent of the fixed times $t$ at which fits to
Eq.~(\ref{eq:prefatt}) have been performed. Likewise, fitted values
of $\tau$ should not depend on $k$. To determine whether this is the
case, we fit $f(k)$ for all possible values of $t$ and observe the
measured values of $\alpha$ in order to check for any systematic
changes with time\footnote{In contrast to the previously considered
case of technology-specific patent-citation data~\citep{higham2017},
the APS-journal-specific citation data are not large enough to enable
an accurate measurement of $f_0$. Motivated by our observation that
$f_0$ is approximately unity for all journals, we henceforth fix
$f_0\equiv 1$. This allows for the accurate measurement of $\alpha$,
whose exact value is much more relevant than that of $f_0$ in
determining the structure and evolution of the citation network.}. We
then perform the same procedure with $A(t)$ across all $k$-value bins
to check for any systematic changes in $\tau$. As illustrated in
Fig.~\ref{fig:parameters}, we indeed find that fitted values simply
fluctuate around a stable mean, thus verifying empirically the
separability of the citation rate $\bar\lambda(k, t)$ in accordance
with Eq.~(\ref{eq:sepaCit}) and with the functional forms for the
aging function and PA kernel given in (\ref{eq:ageing}) and
(\ref{eq:prefatt}), respectively. Table~\ref{parametertable1} lists
the parameter values and their uncertainties that have been extracted
for each individual APS-journal cohort as weighted arithmetic
averages and their 95\% confidence intervals from fitted values such
as those shown in Fig.~\ref{fig:parameters} for PRD, where weights
are the inverses of the variance for each fitted value. The average
for $\alpha$ does not include measurements for times less than $t=2$
years, as there is not enough spread in $k$ for accurate measurement
of this parameter at small times. Results for $A_0$ are calculated
from the fitted values for the product $A_0 f(k)$ at fixed $k$ using
the previously measured value of $\alpha$.

To maximise the accuracy of fit-determined parameters, it turns out
to be useful to adjust the logarithmic bin size in the $k$ dimension
and the time interval $\Delta t$ in order to optimize the measurement
resolution in the variable we are fitting. For example, when fitting
$A(t)$ for the groups of articles in particular fixed-$k$ bins, we
can more accurately represent the data by slightly increasing the
size of the $k$ bins such that more articles are included in each
individual fitting procedure for $\bar\lambda(k, t)$ at fixed $k$,
while at the same time decreasing $\Delta t$ for greater time
resolution. Due to the finite range of empirically available $k$
values, this means there are fewer fixed-$k$ fitting procedures and
thus fewer measurements of $\tau$ and $A_0$; however, we do not
require a large number of measurements to detect any systematic
change in these parameters with $k$. The opposite is true when
fitting $f(k)$. Based on such considerations, we have chosen $\Delta
t = 6$~months when fitting $A(t)$ and $\Delta t  = 1$~year when
fitting $f(k)$.

\begin{table*}[t]
\caption{\label{parametertable1}%
Measured values for the parameters characterising preferential
attachment ($\alpha$) and obsolescence ($\tau$ and $A_0$), extracted
from analysing citations to articles published in APS journals in the
year 2000. For comparison, results obtained from performing the same
analysis on citations to articles published in these journals during
the three-year period 1989--1991 are also given. Uncertainties
represent 95\% confidence intervals. For $\alpha$ ($\tau$) values
labeled with an $\ast$ ($\ast\ast$), the set of averaged values
exhibited a weak residual dependence on time (number of accrued
citations).}
\vspace{0.2cm}
\centering
\begin{tabular}{|c|c|cccccc|}
\hline
Year & Journal & PRA & PRB & PRC & PRD & PRE & PRL \\ \hline
\multirow{3}{*}{2000} &
$\alpha$ & $1.19 \pm 0.08$ & $1.17 \pm 0.12^\ast$ & $1.13 \pm 0.16$
& $1.08 \pm 0.10$ & $1.20 \pm 0.16$ & $1.13 \pm 0.08^\ast$ \\
\cline{2-8}
& $\tau$ [yrs] & $4.86 \pm 0.60$ & $4.37 \pm 0.80^{\ast\ast}$ &
$5.47 \pm 1.55$ & $5.00 \pm 0.80$ & $4.86 \pm 1.18$ & $4.76 \pm
0.87$ \\ \cline{2-8}
& $A_0$ [yrs$^{-1}$] & $0.23\pm 0.03$ & $0.40\pm 0.15$ & $0.27\pm
0.09$ & $0.22\pm 0.04$ & $0.25\pm 0.05$ & $0.36\pm 0.16$ \\ \hline
\multirow{3}{*}{1989-91} &
$\alpha$ & $1.18 \pm 0.09$ & $1.18 \pm 0.08^\ast$ & $1.16 \pm
0.11$ & $1.26 \pm 0.18$ & --- & $1.14 \pm 0.09^\ast$ \\ \cline{2-8}
& $\tau$ [yrs] & $6.33 \pm 0.60$ & $6.73 \pm 0.81^{\ast\ast}$ &
$7.38 \pm 1.44$ & $7.95 \pm 1.60^{\ast\ast}$ & --- & $6.72 \pm 0.85$
\\ \cline{2-8} & $A_0$ [yrs$^{-1}$] & $0.15\pm 0.03$ & $0.13\pm
0.01$ & $0.09\pm 0.04$ & $0.08\pm 0.02$ & --- & $0.14\pm 0.05$
\\ \hline
\end{tabular}
\end{table*}

As is apparent from Fig.~\ref{fig:fits}(a), the exponential form for
$A(t)$ given in Eq.~(\ref{eq:ageing}) turns out to provide a good fit
to the data only for $t \gtrsim 3$~years. We have therefore limited
our fitting of parameters for the aging function to this range. While
other, generally more complicated, functional forms such as shifted
power laws or stretched-exponential functions are able to be fitted
over the whole range of time for our data, the validation of
separability becomes extremely ambiguous with these three-parameter
aging models, because the variances in the measured parameter values
turn out to be very large relative to the magnitude of the parameters
themselves. The full-range fit thus comes at the expense of
meaningful parameter estimations. In contrast, the two-parameter
model of exponential aging enables reliable parameter determination
and accurately represents the data except for a short period after
publication.

It would be very interesting to investigate systematically the
quality of separability and determine the values for parameters in
the aging function and the PA-growth kernel characterizing the
citation dynamics of articles published in different years. However,
cohorts of articles published much earlier than the year 2000 are
generally smaller and have accrued fewer total citations, leading to
larger statistical uncertainties. To improve the statistics and
facilitate at least a glimpse of a basic comparison, we aggregated
and analyzed the citation data for articles published in individual
APS journals in the three-year period 1989--1991. Also for these
earlier-published articles, citations obtained up until 2015 were
included in our analysis. Thus the time range over which aging was
observable for these articles is about ten years longer than for the
year-2000 article cohorts, therefore some caution needs to be
exercised in any direct comparison between the extracted obsolescence
times for articles from the two time periods. The results obtained
from fits to the PA growth kernel from Eq.~(\ref{eq:prefatt}) with
$f_0=1$ and aging function from Eq.~(\ref{eq:ageing}) with $t_0 =
4$~years are also given in Table~\ref{parametertable1}. Short-time
deviations from exponential aging were found to persist over a longer
initial time period for the article cohorts published during
1989--1990 than for the year-2000 cohorts, necessitating the larger
value of $t_0$. Because of this, and the systematically larger
obsolescence time scale found for the earlier-published articles, the
reliable extraction of $\tau$ required including all available data
for citations acquired up until 2015, i.e., for $\sim$25 years after
publication. Note that the observed longer period for deviations from
exponential aging at short times for the earlier published articles
is consistent with their larger $\tau$ values, as we have generally
found these two time scales to be linked~\citep{higham2017}.

\begin{figure}[t]
\centering{\includegraphics[width=0.6\textwidth]{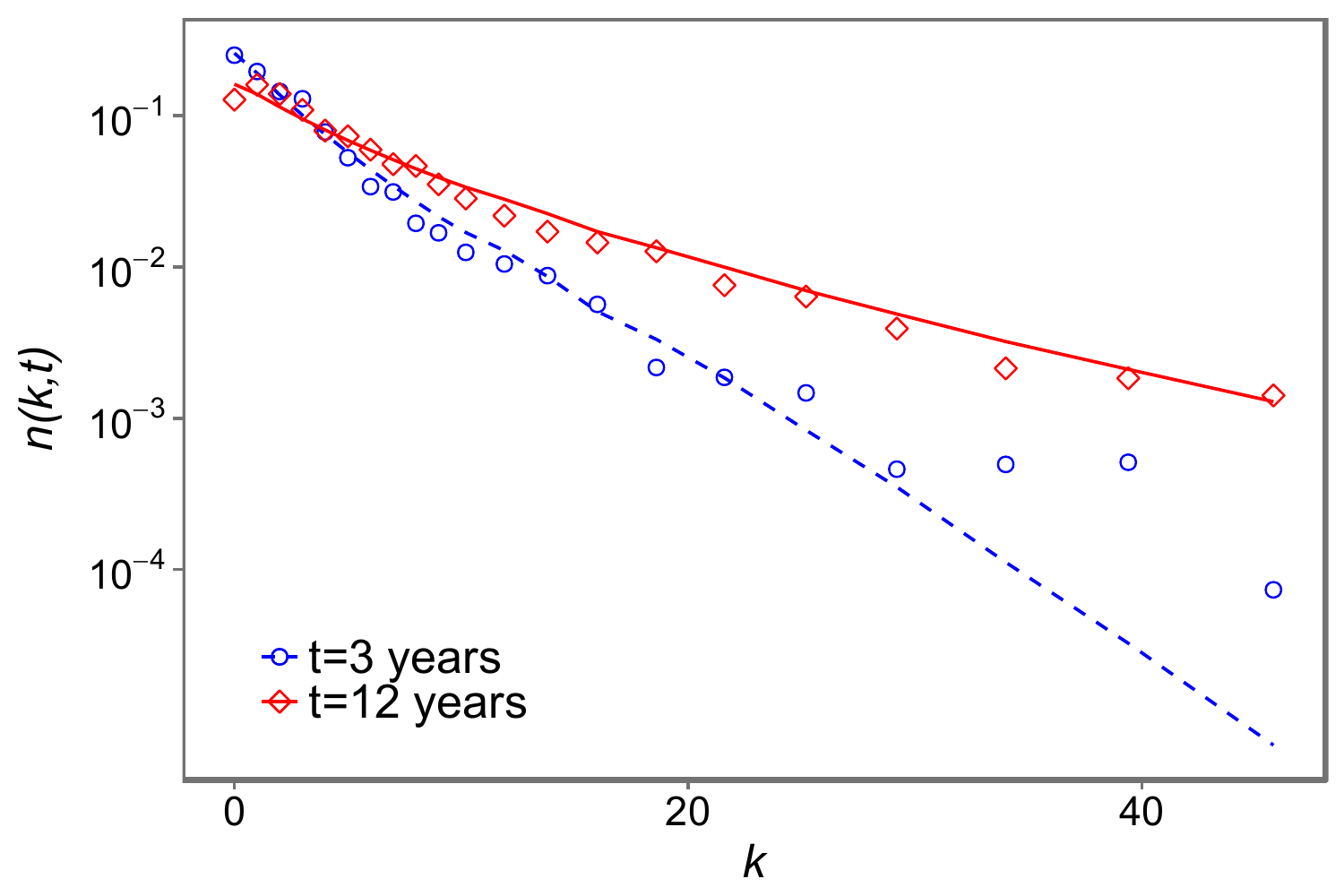}}
\vspace{-0.2cm}
\caption{\label{fig:dispComp}%
Distribution function $n(k, t; T_i)$ for citations to articles
published in PRD in the year 2000, plotted as a function of the
number $k$ of citations for fixed times $t=3\,$years and $t=12
\,$years after publication. Symbols show the empirical data. Curves
were calculated from the general expression Eq.~(\ref{eq:citDistrib})
with parameters given in Table~\ref{parametertable1} for the cohort
of articles published in PRD in the year 2000 and the empirically
observed citation distribution at $t=1\,$year taken as $n_0(k)$. The
form (\ref{eq:expGamma}) was used with $\Gamma_0 = 0.62$. The
disagreement between theory and data seen for large $k$ at small
$t$ is due to the systematic deviation from exponential aging
occuring at short times [see Fig.~\ref{fig:fits}(a)].}
\end{figure}

A citation rate of the separable form Eq.~(\ref{eq:sepaCit}), with
an aging function $A(t;T_i)$ that can, in principle, depend also
on the publication time $T_i$, gives rise to a distribution function
$n(k, t; T_i)$ for citations to articles published at the same time
that can be expressed most generally as~\citep{higham2017}
\begin{equation}\label{eq:citDistrib}
n(k, t; T_i) = n_0(k) \left[\gamma(t; T_i)\right]^{f(k)} +
\sum_{l=0}^{k-1} n_0(l) \left( \prod_{m=l}^{k-1} f(m) \right)
\sum_{q=l}^k \frac{\left[\gamma(t; T_i)\right]^{f(q)}}{\prod_{m=l
\atop m\ne q}^k \left[f(m) - f(q) \right]} \quad .
\end{equation}
Here
\begin{equation}\label{eq:gamma}
\gamma(t; T_i) = \exp\left\{-\int_0^{t} dt'\, A(t'; T_i)\right\}
\quad ,
\end{equation}
and $n_0(k) \equiv n(k, 0; T_i)$ encodes a fully general initial
condition for the distribution. Figure~\ref{fig:dispComp} shows a
comparison between empirical data for $n(k, t; T_i)$ for the cohort
of articles published in PRD in the year 2000 and the theoretical
prediction obtained from Eq.~(\ref{eq:citDistrib}) with the parameter
values for aging and PA-driven growth given in 
Table~\ref{parametertable1} for this cohort. We used the expression
\begin{equation}\label{eq:expGamma}
\gamma(t; T_i) = \Gamma_0 \exp\left\{ \tau A_0 \left[ \exp \left( -
\frac{t}{\tau} \right) - 1 \right] \right\} \quad ,
\end{equation}
which is the result obtained from Eq.~(\ref{eq:gamma}) with
Eq.~(\ref{eq:ageing}) as the aging function, rescaled by the
cohort-specific factor $\Gamma_0=0.62$ to account for the observed
short-term deviations from exponential aging~\citep{higham2017}.
To minimize the impact of having arbitrarily fixed $f_0=1$ in our
fitting procedure, we took the empirically observed distribution
of citations at $t=1\,$year as $n_0(k)$. The agreement between theory
and data is excellent, except for large $k$ at small $t$ where the
influence of deviations from exponential aging at short times cannot
be properly quantified by our model based on the parameter
$\Gamma_0$~\citep{higham2017}.

Knowledge of $n(k, t; T_i)$ in principle allows to also derive the
distribution function $P(k|T_<, T_>)$ for citations in the aggregated
article network comprising cohorts with $T_i\in [ T_<, T_>]$;
\begin{equation}
P(k; T_<, T_>) = \frac{1}{T_>-T_<} \int_{T_<}^{T_>} d T \, n(k, T_>
- T; T) \quad .
\end{equation}
Citation distributions for large collections of articles whose
publication times span long time intervals have been the subject of
intense recent interest~\citep{redner2005,stringer2010,radicchi2011,
waltman2012,subelj2017,yin2017,sheridan2017}. One particular question
such studies have aimed to answer is how the empirical distributions
compare with the form of stationary citation distributions $P(k)
= \lim_{T_>\to\infty} P(k; T_<, T_>)$ arising within the framework of
relevant network models. Analyzing this issue for articles within
separate research fields could be an interesting direction for future
research~\citep{stringer2010}.

\section{Discussion}

Gaining a full understanding of the dynamics of citation accrual by
scientific articles and patents has been hampered by the need to
account for various closely entangled, and sometimes mutually
counteracting, influences. On a basic qualitative level, it can be
surmised that diffusion of knowledge drives the accumulation of
citations by articles or patents that report useful new
information~\citep{jaffe1999,bornmann2008}. The ability to accurately
model, and potentially optimise, the dynamics of knowledge diffusion
should be facilitated by the increased availability of high-quality
citation data, if only the additional mechanisms affecting real-world
citing behaviour could be reliably identified and accounted for. The
results of our present study constitute a step in this direction.

One factor that is as intrinsic to the scientific (or invention)
ecosystem as knowledge diffusion is the process of obsolescence,
i.e., the tendency for previously codified information and methods to
become less relevant over time for continued progress of knowledge
generation. Conceptually, obsolescence can be understood as an aging
process. However, both in the real world and in idealised
network-model descriptions, any existing intrinsic aging dynamics can
be masked by exogenous influences that effectively contribute to
aging or counterbalance it. For example, in the generic network model
applied to citing behaviour~\citep{dSPrice1976,barabasi1999} where a
new node is added at each time step $T$ and distributes a number of
citations to existing nodes according to a PA mechanism, the
linear-in-time growth of the overall network induces a purely
structural aging process. The associated aging function is given
by~\citep{albert2002} $A(t; T_i) = 1/[2(T_i + t)]$, where $T_i$ is
the time at which the node has been added (corresponding to the
publication time for an article or patent). In the real world, the
rapid increase over time in the overall number of published articles
has similarly been perceived as a structural cause of
aging~\citep{parolo2015}. On the other hand, an increase in the rate
of production for articles (patents) that cite relevant prior
knowledge can boost the citation rate for older articles (patents).
In fact, the combination of increased publication activity and an
on-average increased number of citations made per article (patent)
has been seen to cause a citation inflation that partially
compensates aging effects~\citep{hall2002,wang2013,pan2016}.

Our present approach is designed to carefully disentangle the
mechanism of PA-driven citation accumulation from the effects of
aging and inflation. The litmus test for having achieved this goal is
provided by the absence of residual time dependences in PA-related
parameters, especially the exponent $\alpha$, that have been
extracted from fits of the empirical citation rate. See
Fig.~\ref{fig:parameters}(b). Furthermore, the parameters governing
obsolescence-related aging should be found to be independent of the
number $k$ of citations, as is indeed the case
[Fig.~\ref{fig:parameters}(a)]. To be able to demonstrate this clear
separation, we needed to focus our analysis on specific research
fields in physics, as defined by the scope of individual APS
journals, and account for citation inflation by the journal-specific
scaling factor $c_J$. In contrast, previous studies that did not
separate articles by research fields and did not account for citation
inflation found significant monotonous increases over time in the
extracted values of $\alpha$~\citep{golosovsky2012}. We observe some
deviations from full separability of the empirical citation rate into
independent aging and PA parts [Eq.~(\ref{eq:sepaCit})] for the cases
of PRB (see results presented in \ref{sec:appB}) and PRL, which are
both journals that publish articles from a much broader and more
heterogeneous range of physics subfields, and even from neighboring
disciplines such as chemistry and mathematics, than the other four
journals. Our results suggest that the dynamics of knowledge
diffusion and intrinsic obsolescence of knowledge is
research-field-specific.

We identify the obsolescence-related aging function to be an
exponential function of time since publication in the long term, as
expressed in Eq.~(\ref{eq:ageing}). This observation broadly agrees
with recent studies of larger scientific-article
cohorts~\citep{parolo2015,pan2016}, although we find a slightly
shorter value for the obsolescence time scale $\tau$ in our
research-field-specific analysis and also observe some variation
of this parameter between the different fields. In particular, the
subfields associated with PRC and PRD appear to be more slowly
changing than those covered by the other three specialised journals
that turn out to have a similarly short obsolescence time as the
multidisciplinary journal PRL. While not fully conclusive because of
the overall scale of uncertainties in the extracted $\tau$ values,
this observation is consistent with expectations based on known
characteristics of the research fields covered by PRC and PRD,
especially a dependence on the long-term development of large-scale
equipment run by very large consortia of researchers.

Deviations from exponential-in-time aging occur at short times $t
\lesssim 3$~years [see Fig.~\ref{fig:fits}(a)], with more citations
getting accumulated per unit time than expected from an extrapolation
of the long-term exponential-aging trend. A similar excess of
citations arriving in a short time period after publications was also
observed for patents~\citep{higham2017}. The apparent universality of
this short-term enhancement suggests the existence of a common origin
related to knowledge-flow dynamics, which should be clarified by
systematic further studies of the citation-number dependence of the
excess number of citations (generally deviations from exponential
aging are found to be smaller in cohorts of more highly cited
articles or patents), as well as possible systematic variations
across different research fields or technology categories.
Interestingly, the relative magnitude of the short-term deviation
from exponential aging exhibited by low to medium-cited
scientific-article cohorts is consistently observed to be larger, by
at least a factor of two, than for the comparable cohorts of patents.

The fits of the PA-mechanism growth kernel given in
Eq.~(\ref{eq:prefatt}) with fixed $f_0\equiv 1$ to the data yield
values of the exponent $\alpha$ that vary moderately across the
different APS-journal cohorts and are generally consistent with a
superlinear dependence on $k$ ($1.0 < \alpha \lesssim 1.2$). Due to
the relatively small data set as compared, e.g., with our previous
patent-citation study~\citep{higham2017}, the functional form of
Eq.~(\ref{eq:prefatt}) yields only marginally better fits than the
alternative form given in Eq.~(\ref{eq:altPA}) that has also been
applied previously to scientific-article citation
data~\citep{golosovsky2012} but was ruled out for
patents~\citep{higham2017}. See the more detailed discussion in
\ref{sec:appA} and results presented in Fig.~\ref{fig:AltModel} and
Table~\ref{parametertable2}. The values for $\alpha$ obtained using
the two alternative PA-kernel forms are essentially the same.

The aggregation of citation data for articles published in individual
APS journals over the three-year period 1989--1991 made them amenable
to the same type of analysis performed on the year-2000 article
cohorts. Interestingly, we found the same qualitative features and
even some of the same quantitative results for the citation dynamics
of the earlier-published articles as for those from the year 2000. In
particular, PA-driven growth of citations is clearly and robustly
exhibited, with values for the exponent $\alpha$ essentially the same
as for the respective year-2000 APS-journal cohorts. Detailed results
are given in Table~\ref{parametertable1}. Aging was again found to be
described by an exponential function of time in the long run also for
the cohorts of articles published during 1989--1991, but with
prominent deviations from exponential aging at short times that
persist over a longer initial time period ($\sim$4 years) than for
the cohorts of articles from 2000. As a result, reliable extraction
of the obsolescence time $\tau$ associated with the long-term
exponential-aging behavior required to fit all the available data
upto the end of the $\sim$25-year citation history of the 1989--1991
articles. Their thus-obtained values for $\tau$ are consistently
larger, by about $2$~years, than the $\tau$ values we found for the
articles published in the same journal in 2000 using the data from
their shorter (only $\sim$15-year-long) citation history. At the same
time, the obsolescence times of the older articles show the same
trends for variation across journals (i.e., PRC and PRD again
appearing to be associated with more slowly changing subfields). The
juxtaposition of parameters describing PA and aging of citations to
articles published ten years apart already provides an interesting
snapshot of temporal variations in journal-article citation dynamics.
There is scope to perform similarly suitable aggregations over
multi-year periods to analyze even older APS-journal-article cohorts.
Alternatively, larger-scale studies of potential time variations of
citation-dynamics parameters within different physics subfields would
require going beyond the APS-journal data set, thus creating the need
to define association with a subfield either based on a broadly
adopted classification scheme such as PACS or via a more fine-grained
version of a previously employed topic-specific analysis of citation
patters~\citep{sinatra2015}.

\section{Conclusions}\label{sec:concl}

We have analyzed the time evolution of citation data for articles
published in six different APS journals in 2000 to gain insight into
research-field-specific characteristics of knowledge-flow dynamics.
Unlike previous studies, we have accounted for citation inflation
arising from temporal variations in the rate of publication of
articles in the individual journals by a normalization factor. We
demonstrate separability of the empirical citation rate for most
journal-article cohorts into a purely citation-number-dependent part
that reflects a preferential-attachment-driven growth mechanism and a
purely time-dependent aging part that is an exponential function of
time in the long term. Deviations from full separability that are
observed for PRB and, to a lesser extent, PRL are smaller in
magnitude than in previous studies where articles were not separated
by research fields~\citep{golosovsky2012}, suggesting that such
deviations are likely caused by the underlying heterogeneity of
scientific communities publishing in these two journals.

As exogenously caused variations in citability (`inflation') are
accounted for within our approach, the observed characteristics of
the aging function should be dominated by the dynamics of
obsolescence for knowledge within the specific research fields. This
provides a window into the dynamics of scientific progress within
these fields, as the time scale for obsolescence is commonly
associated with the rate at which the knowledge frontier advances. In
particular, we were able to identify more slowly changing research
fields (those associated with PRC and PRD) compared to the rest that
have a similar obsolescence time as the multidisciplinary journal
PRL. Furthermore, application of our analysis to cohorts of
APS-journal articles published in the period 1989-1991 revealed their
obsolescence times to be about 2 years longer compared to the
year-2000 cohorts from the same journal, while following the same
basic trend in the variation across different journal cohorts.
However, the fact that deviations from the well-understood
exponential obsolescence behavior occur over a longer initial time
period for the articles published during 1989-1991 --- requiring us
to include the data from the period $\sim$15-25 years after their
publication to reliably extract $\tau$ --- makes a direct comparison
with the articles published in 2000 somewhat difficult, as no
citation data beyond $\sim$15 years after publication are available
for the latter. More detailed analysis is needed to exclude potential
other systematic influences that affect the obsolescence of articles
15-25 years after their publication. If the observation of faster
obsolescence for the younger article cohorts were to be robustly
substantiated, then it could signify an overall accelerating pace at
which science advances in all subfields of physics. Such an
acceleration might, for example, be rooted in changing patterns of
exploration (focus on broad and general \textit{versus\/} deep and
specific), but evolving citation practices or reading habits can also
affect the obsolescence time scale~\citep{egghe2000}. 

Although the obsolescence-induced aging is accurately described by
an exponential function for intermediate to long times after
publication, an excess of citations above the extrapolated
exponential behavior occurs within the first 2-3 (4-5) years after
publication for the articles from 2000 (1989--1991), with stronger
deviations occurring for the article cohorts with smaller number of
accrued citations. More systematic study is needed to determine the
origin of these deviations, but their particular features point to
the existence of a special knowledge-propagation mechanism that is
effective at short times.

Even though the mechanisms and motivations determining citing
behavior of academics and inventors have been identified to be quite
different \citep{bornmann2008,cotropia2013,jaffe2017}, the results of
our present study turn out to be strikingly similar, both
qualitatively and quantitatively, to those found previously in an
analysis of the inflation-adjusted citation dynamics for patents
granted in 1998 within specific technology
categories~\citep{higham2017}. In particular, the values of the
exponent $\alpha$ characterizing preferential-attachment-type growth
vary over the same basic range of magnitude between the different
article and patent cohorts. The obsolescence time is observed to be
just slightly longer for the patents compared to that of the
APS-journal articles, whereas the magnitude of the citation-rate
enhancement over the extrapolated exponential-aging behavior in the
short term is systematically larger for the cohorts of articles than
for the comparable patent cohorts. Our ability to provide a more
detailed comparison between the citations dynamics of patents and
scientific articles is hampered by the fact that the accuracy of
parameter values extracted in the present study was more limited due
to the smaller size (by roughly an order of magnitude both in total
numbers of accrued citations and in total numbers of citable items)
of the article cohorts in comparison with the patent cohorts. Further
studies of research-field specific trends in article-citation
dynamics will need to utilize larger data sets that have reliably
tagged outputs from different research fields. Demonstrating the
separability of the appropriately inflation-adjusted empirical
citation rate for these larger cohorts into a purely
citation-number-dependent growth part and an obsolescence-induced
aging part will be a crucial first step to obtain, and meaningfully
compare, relevant parameters.

Questions to be addressed by future studies include the relevance
of memory effects in the citation dynamics that can cause deviations
from preferential-attachment-type growth. This aim will also require
use of larger data sets as, e.g., autocorrelations were previously
observed to be significant only for the very highly cited ones among
scientific articles~\citep{golosovsky2012}. We do not expect such
effects to be relevant for our present analysis where articles were
excluded once their citation count reached the 99th percentile for
their respective journal-article cohort. Another interesting issue
that could be explored concerns the functional form of the
steady-state distribution of citations to articles within a given
specialized research field~\citep{stringer2010}. Whether and how
citation inflation is accounted for may crucially influence the
observed properties of such distributions~\citep{radicchi2011,
waltman2012,yin2017,subelj2017}. Further systematic investigation of
this question could inform ongoing discussions about the consistency
of PA-driven growth models with empirically observed static
properties of citation networks~\citep{golosovsky2017a,sheridan2017}.

\appendix
\renewcommand{\thefigure}{\arabic{figure}}
\renewcommand{\thetable}{\arabic{table}}
\renewcommand{\thesection}{Appendix~\Alph{section}}

\section{Results from fitting an alternative functional form for the
preferential-attachment kernel}\label{sec:appA}

\begin{table*}
\caption{\label{parametertable2}%
Measured values for $\alpha$, and those derived for $A_0$, obtained
from fitting the functional form $f(k) = (k + 1)^\alpha$ for the PA
kernel to the citation data for articles published in APS journals in
the year 2000. For the $\alpha$ values labeled with an $\ast$, the
set of averaged values exhibited a weak residual dependence on time.
Note the only slightly larger uncertainties for the $\alpha$ values
in comparison with results given in Table~\ref{parametertable1}.}
\vspace{0.2cm}
\centering
\begin{tabular}{|c|cccccc|}
\hline
Journal & PRA & PRB & PRC & PRD & PRE & PRL \\ \hline
$\alpha$ & $1.21 \pm 0.09$ & $1.19 \pm 0.13^\ast$  & $1.15 \pm 0.19$
& $1.09 \pm 0.11$ & $1.22 \pm 0.19$ & $1.14 \pm 0.10^\ast$ \\ \hline
$A_0$ [years$^{-1}$] & $0.22\pm 0.04$ & $0.38\pm 0.18$ & $0.26\pm
0.09$ & $0.22\pm 0.03$ & $0.23\pm 0.07$ & $0.34\pm 0.17$ \\ \hline
\end{tabular}
\end{table*}

\begin{figure}[t]
\centering{\includegraphics[width=0.48\textwidth]{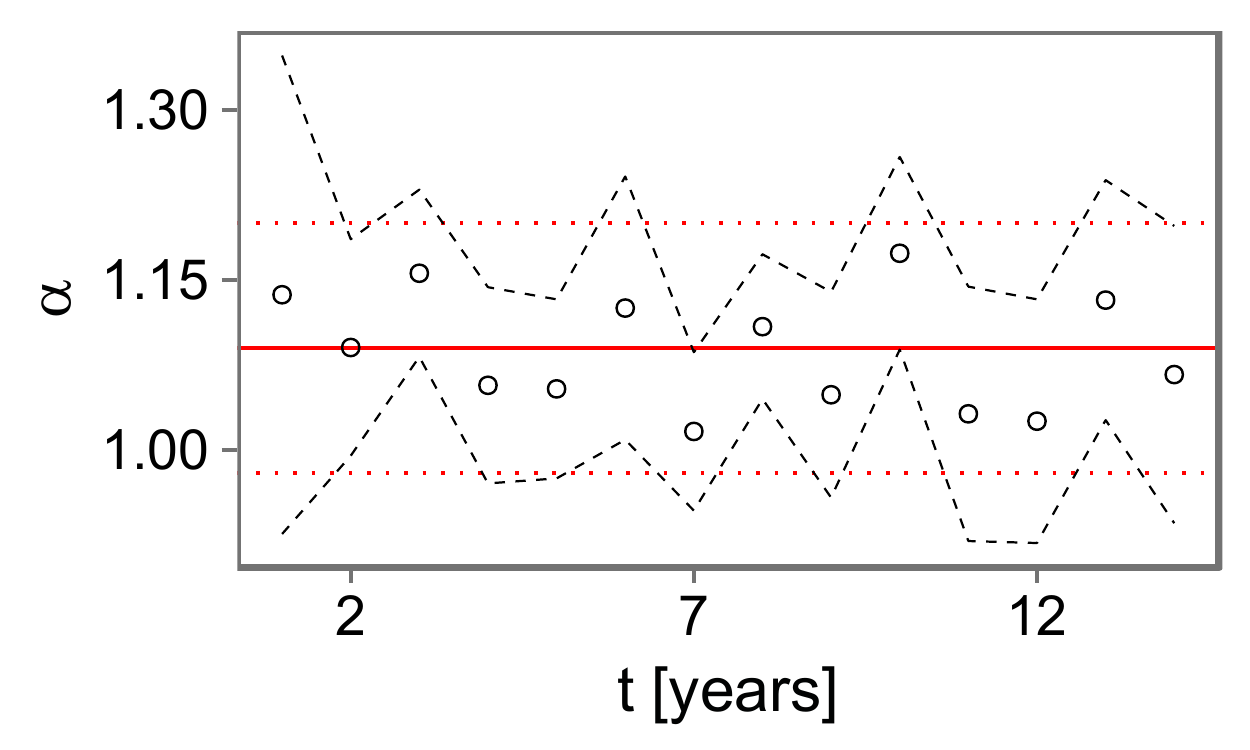}}
\vspace{-0.4cm}
\caption{\label{fig:AltModel}%
Exponent $\alpha$ extracted from fitting a PA growth kernel of the
form $f(k) = (k + 1)^\alpha$ to the empirical citation rate for
articles published in PRD in 2000. Circles are the fit values, the
solid line is their weighted average, and black dashed (red dotted)
curves indicate 95\% confidence intervals for the fit values (the
weighted average). Note the extremely small differences with the
results shown in Fig.~\ref{fig:fits}(b).}
\end{figure}

A number of functional forms have been utilised to characterise
superlinear preferential attachment, most of which converge to
$k^\alpha$ in the large-$k$ limit. We have adopted one such form,
given in Eq.~(\ref{eq:prefatt}), to model preferential attachment in
this work. An alternate form of preferential attachment incorporates
a constant shift into the argument,
\begin{equation}\label{eq:altPA}
f(k)=(k+k_0)^\alpha \quad .
\end{equation}
We have fitted also this functional form to the empirical citation
rates in order to test whether this provides a better fit, using
fixed $k_0=1$ in analogy to our methodology in the main body of this
work. The results obtained from the alternative-fit analysis are
summarised in Table~\ref{parametertable2} and
Fig.~\ref{fig:AltModel}. While uncertainties in the measured values
of $\alpha$ for the various journals turn out to be slightly larger
than the values obtained by fitting to Eq.~(\ref{eq:prefatt}) (see
results presented in Table~\ref{parametertable1}), there is little
quantitative difference between the measurements of $\alpha$ from
the two models.

\section{Observed deviations from full separability of the
empirical citation rate: Case of PRB}\label{sec:appB}

\begin{figure}[t]
\includegraphics[width=0.48\textwidth]{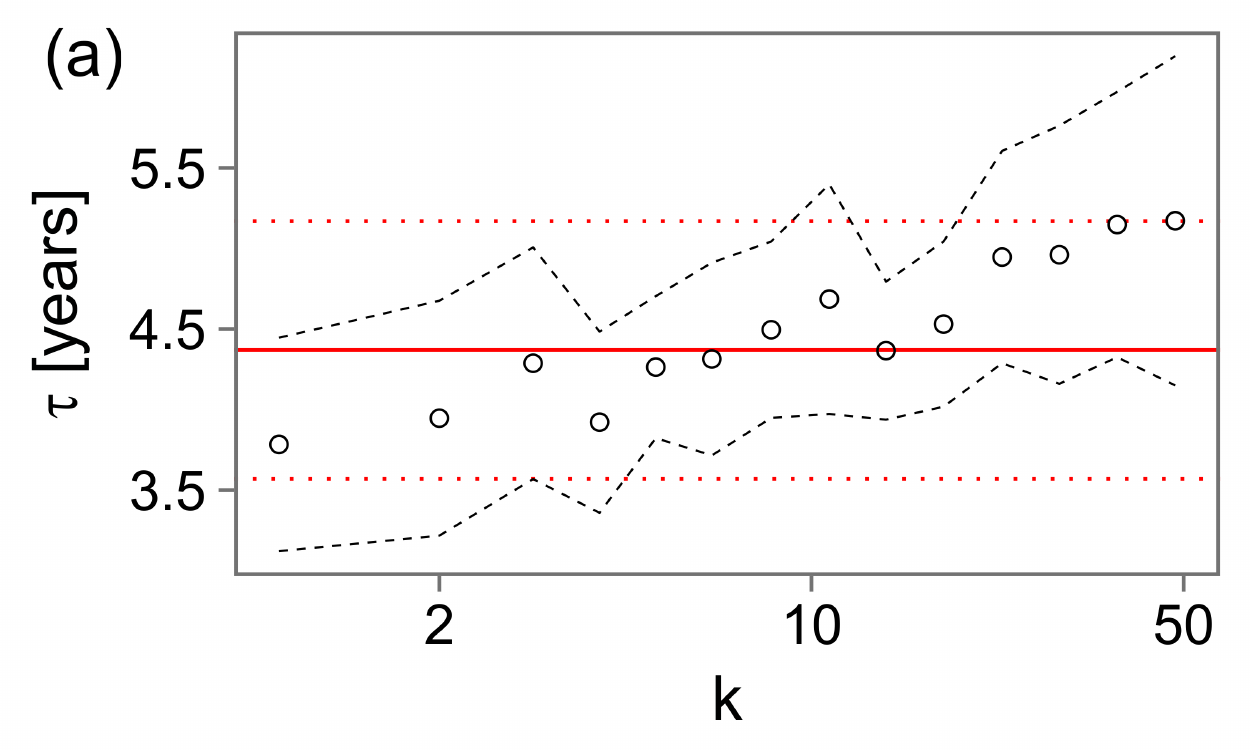}\hfill
\includegraphics[width=0.48\textwidth]{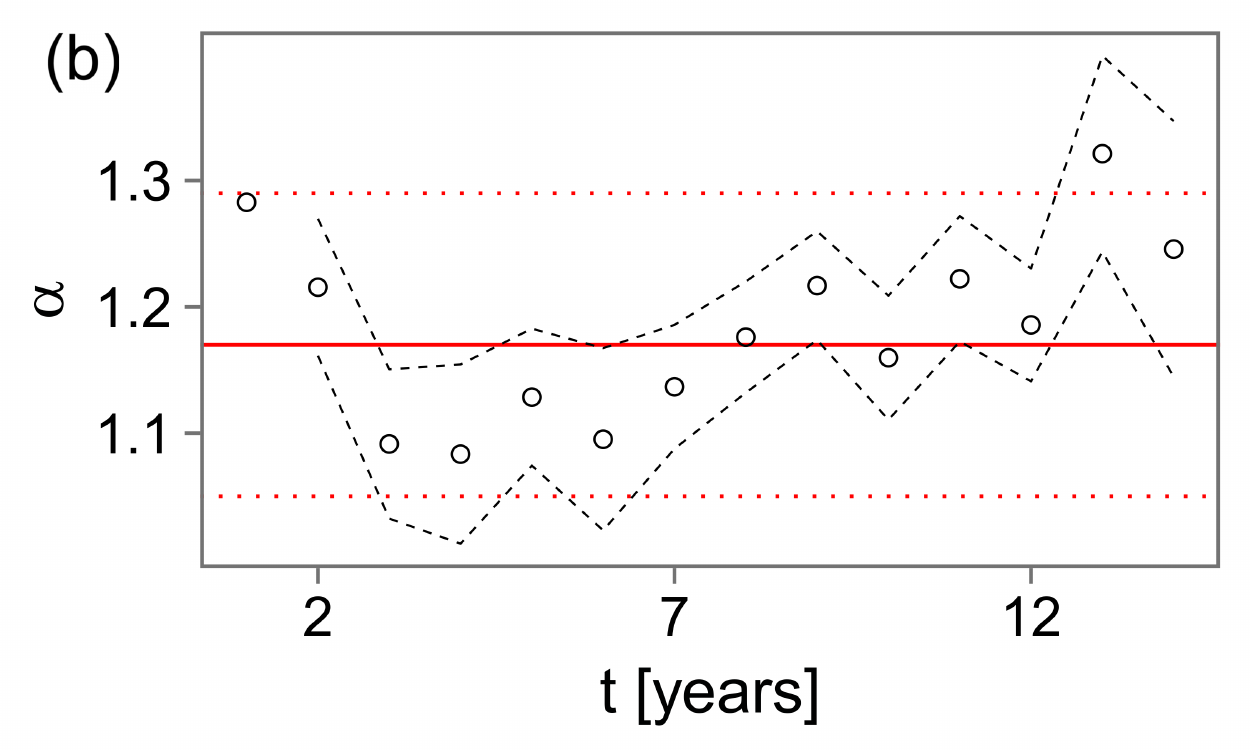}
\vspace{-0.4cm}
\caption{\label{fig:deviation}%
Illustration of how deviations from separability of the empirical
citation rate are manifested. Data shown pertain to articles
published in PRB in the year 2000. Circles are the fitted parameter
values, solid lines indicate their weighted averages, and the black
dashed (red dotted) curves show the 95\% confidence intervals for
fit-parameter values (their weighted averages). (a)~Values of the
obsolescence time $\tau$ extracted from fits of the empirical
citation rate to the functional form $A(t)$ from
Eq.~(\ref{eq:ageing}) for different fixed $k$. In contrast to the
case shown in Fig.~\ref{fig:parameters}(a), a systematic trend for
$\tau$ values to increase as a function of $k$ is exhibited here.
(b)~Values for the exponent $\alpha$ derived from fits of the
empirical citation rate to the form (\ref{eq:prefatt}) for the PA
growth kernel $f(k)$, assuming $f_0=1$. In comparison with
Fig.~\ref{fig:parameters}(b), $\alpha$ shows a systematic dependence
on $t$.}
\end{figure}

The separation of the empirical citation rate into independent
factors describing long-term exponential aging and PA-driven citation
accumulation, as expressed in Eq.~(\ref{eq:sepaCit}), was
demonstrated for (most of) the APS-journal-article cohorts by
observing relevant fit parameters for the aging function (the PA
growth kernel) to be independent of the variable $k$ ($t$). The
general quality of the demonstrated separation is illustrated in
Fig.~\ref{fig:parameters} using the data for articles published in
PRD. A deviation from separability was observed for PRL where the
extracted values for the exponent $\alpha$ exhibited a trend to
increase as a function of $t$, varying between 1.07 and 1.17 over
our study's 15-year period. However, the most drastic violation of
separability occurred for PRB where both $\alpha$ and $\tau$ showed
residual dependences on $t$ and $k$, respectively.

Figure~\ref{fig:deviation} shows the results for PRB. The observed
trend of increasing $\alpha$ as a function of $t$ is slower than, but
still of roughly the same order of magnitude as, in studies where
articles were not disaggregated by research field and inflation was
not accounted for~\citep{golosovsky2012}. The increasing advantage of
more highly cited articles to attract further citations at a higher
rate could reflect the greater importance of autocorrelations in the
citation dynamics of PRB and PRL articles. Alternatively, a greater
heterogeneity in terms of research field and stronger
multidisciplinary influences from fields outside physics that
characterize both PRB and PRL could be the cause. Support for this
conclusion is also provided by the results of a related
patent-citation study~\citep{higham2017} where deviations from
separability of the empirical citation rate also occurred for the
more heterogeneous technology categories.

%
%

\end{document}